\documentclass[conference]{IEEEtran}
\IEEEoverridecommandlockouts
\usepackage{cite}
\usepackage{amsmath,amssymb,amsfonts}
\usepackage{algorithmic}
\usepackage{graphicx}
\usepackage{textcomp}
\usepackage{xcolor}
\usepackage{hyperref}
\usepackage{array,multirow}

\usepackage{makecell}

\def\BibTeX{{\rm B\kern-.05em{\sc i\kern-.025em b}\kern-.08em
    T\kern-.1667em\lower.7ex\hbox{E}\kern-.125emX}}
\begin{document}

\title{Deep Learning Methods for Retinal Blood Vessel Segmentation: Evaluation on Images with Retinopathy of Prematurity\\
}

\author{\IEEEauthorblockN{1\textsuperscript{st} Gorana Gojić}
\IEEEauthorblockA{\textit{Faculty of Technical Sciences} \\
\textit{University of Novi Sad}\\
Novi Sad, Serbia \\
gorana.gojic@uns.ac.rs}
\and
\IEEEauthorblockN{2\textsuperscript{nd} Veljko Petrović}
\IEEEauthorblockA{\textit{Faculty of Technical Sciences} \\
\textit{University of Novi Sad}\\
Novi Sad, Serbia \\
pveljko@uns.ac.rs}
\and
\IEEEauthorblockN{3\textsuperscript{rd} Radovan Turović}
\IEEEauthorblockA{\textit{Faculty of Technical Sciences} \\
\textit{University of Novi Sad}\\
Novi Sad, Serbia \\
radovan.turovic@uns.ac.rs}
\and
\IEEEauthorblockN{4\textsuperscript{th} Dinu Dragan}
\IEEEauthorblockA{\textit{Faculty of Technical Sciences} \\
\textit{University of Novi Sad}\\
Novi Sad, Serbia \\
dinud@uns.ac.rs}
\and
\IEEEauthorblockN{5\textsuperscript{th} Ana Oros}
\IEEEauthorblockA{\textit{Dept. of Retinopathy of Prematurity and Retinal Development}\\
\textit{Institute of Neonatology}\\
Belgrade, Serbia \\
annaoros03@yahoo.com}
\and
\IEEEauthorblockN{6\textsuperscript{th} Dušan Gajić}
\IEEEauthorblockA{\textit{Faculty of Technical Sciences} \\
\textit{University of Novi Sad}\\
Novi Sad, Serbia \\
dusan.gajic@uns.ac.rs}
\and
\IEEEauthorblockN{7\textsuperscript{th} Nebojša Horvat}
\IEEEauthorblockA{\textit{Faculty of Technical Sciences} \\
\textit{University of Novi Sad}\\
Novi Sad, Serbia \\
horva.n@uns.ac.rs}
}

\maketitle

\begin{abstract}
Automatic blood vessel segmentation from retinal images plays an important role in the diagnosis of many systemic and eye diseases, including retinopathy of prematurity. Current state-of-the-art research in blood vessel segmentation from retinal images is based on convolutional neural networks. The solutions proposed so far are trained and tested on images from few available retinal blood vessel segmentation datasets, which might limit their performance when given an image with retinopathy of prematurity signs. In this paper we evaluate performance of three high-performing convolutional neural networks for retinal blood vessel segmentation in context of blood vessel segmentation on retinopathy of prematurity retinal images. The main motive behind the study is to test if existing public datasets suffice to develop a high performing predictor that could assist an ophthalmologist in retinopathy of prematurity diagnosis. To do so, we create a dataset consisting solely of retinopathy of prematurity images with retinal blood vessel annotations manually labeled by two observers, where one is the ophthalmologist experienced in retinopathy of prematurity treatment. Experimental results show that all three solutions have difficulties in detecting the retinal blood vessels of infants due to a lower contrast compared to images from public datasets as demonstrated by significant drop in classification sensitivity. All three solutions segment alongside retinal also choroidal blood vessels which are not used to diagnose retinopathy of prematurity, but instead represent noise and are confused with retinal blood vessels. By visual and numerical observations, we observe that existing solutions for retinal blood vessel segmentation need improvement toward more detailed datasets or deeper models in order to assist the ophthalmologist in retinopathy of prematurity diagnosis. 
\end{abstract}


\section{Introduction}
\begin{figure}[!t]
	\centering
	\includegraphics[scale=0.35]{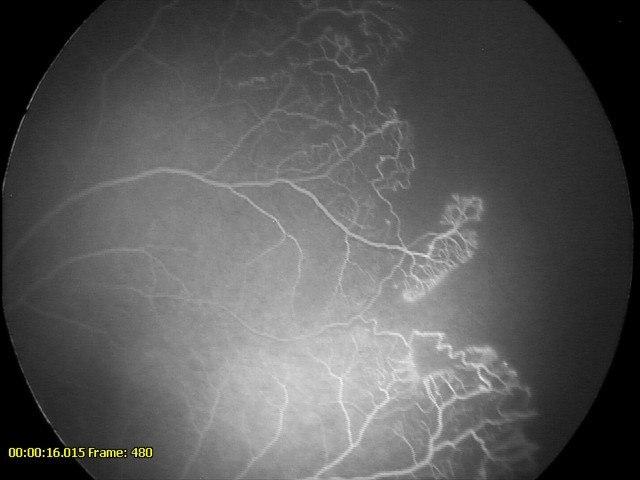}
	\caption{Eye retinal image showing excessive blood vessel growth and appearance of closed blood vessel paths.}
	\label{fig:angiography}
\end{figure}

Retinopathy of prematurity (ROP) \cite{c15} is an eye disease affecting prematurely born infants. It manifests itself through anomalies in the retina which is the back part of the eye. Abnormal chemical changes in the eye result in interruption of normal blood vessel development causing excessive vessel growth in affected areas and appearance of vessel loops as can be seen in Fig. \ref{fig:angiography}. Since the normal vessel growth is suspended, parts of the retina stay avascularized, and if not treated, detach over time and finally lead to blindness. 

To effectively treat ROP, and prevent the possibility of disease developing into blindness, it is of great importance to diagnose it in early stages of development. Prematurely born infants fulfilling certain criteria are screened for ROP to check for the retinal blood vessel tortuosity and dilatation, which are initial ROP symptoms. These changes can be observed in digital images captured by specialized wide-field fundus cameras such as RetCam. In the first column of Fig. \ref{fig:dataset} we show retinal images captured with a RetCam3 camera. However, ROP diagnosis from an image can be challenging due to issues such as: (1) low contrast between blood vessels and retina, (2) variable illumination in image due to wide-field view, (3) relatively low image resolution and (4) appearance of choroidal blood vessels which might be visible on image but do not play part in ROP diagnosis \cite{c16}. 

To assist ophthalmologists in disease diagnosis from digital retinal images, many general, fully-automated blood vessel segmentation solutions have been developed. Since the subset of ROP symptoms manifests itself through blood vessel changes, it can be expected that these solutions could assist an ophthalmologist in disease diagnosis. Current state-of-the-art solutions for retinal blood vessel segmentation are based on convolutional neural networks (CNNs) \cite{c5,c6,c7,c8,c11,c12,c14}. It is reported that these solutions achieve outstanding segmentation results confirmed both visually and numerically. However, it is known that CNNs performance depends on the data used for training. All high performing solutions in this field are trained using publicly available retinal segmentation datasets counting up to 40 annotated retinal images of adults or school children. We observe that images from publicly available datasets are visually different and apparently easier to segment compared to the retinal images of infants showing signs of ROP. Some differences between images from those datasets and ROP images concern: (1) a blood vessel network which is fully developed in images from publicly available datasets occasionally showing signs of degradation, while ROP images show just partially developed blood network, (2) blood vessels that are thicker compared to infant blood vessels, (3) a contrast between blood vessels and retina which is usually stronger, and (4) the illumination distribution which is uneven in wide-field ROP images opposite to narrow-field images from the public datasets. 

Although the data class remains the same, existing solutions trained on public datasets might degrade in performance when tested on ROP images. It is our intention to measure the impact on the network performance when given a ROP image and to determine if the existing solutions can be used as-is to assist in ROP diagnosis. Since these solutions perform well when tested on publicly available datasets, poor results in this study could indicate that existing publicly available datasets do not suffice to develop high-performing CNN that can be used for blood vessel segmentation on ROP images. To test the hypothesis, we choose three recently developed, high-performing CNNs proposed by Li et al. \cite{c14}, Oliveira et al \cite{c6}, and Zhuang \cite{c13}. We feed pretrained models with 9 ROP images provided by the Institute of Neonatology in Belgrade. The blood vessels on the images are manually labeled by two observers, including one ROP expert with more than 20 years of experience in ROP treatment, to create ground truth labels which are used to evaluate the selected networks.

The rest of the paper is organized as follows: in Section \ref{sec:related-work} we give an overview of the datasets commonly used to train CNNs for blood vessel segmentation. Also, we introduce the reader to the recent progress in the development of CNNs for blood vessel segmentation from color retinal images. Section \ref{sec:methodology} discusses the details of the methodology used, including the preparation of a ROP dataset, the experimental protocol, and used evaluation metrics. Obtained results are presented and discussed in Section \ref{sec:results-and-discussion}. The final section offers the main conclusions as well as directions for the future work.

\section{Related Work}\label{sec:related-work}
\begin{figure*}
    \includegraphics[width=\textwidth]{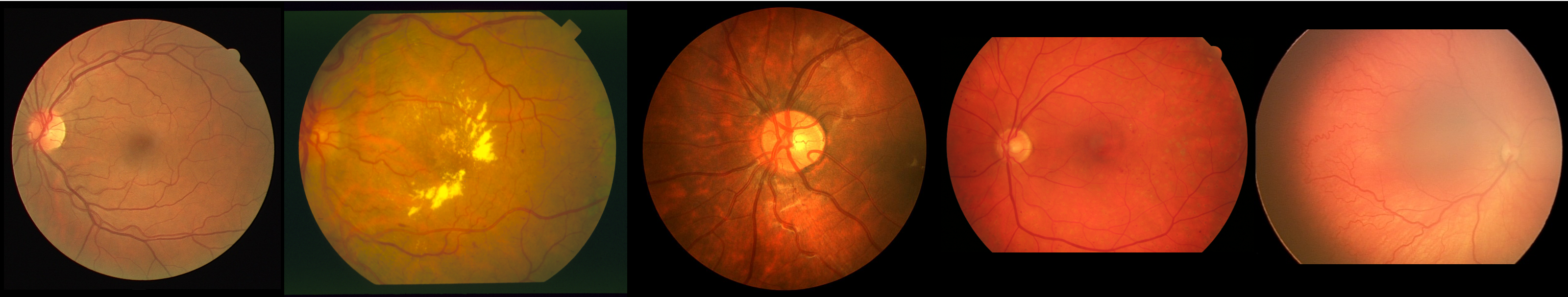}
    \caption{Example images from retinal vessel segmentation datasets. From left to right are representatives of: (1) DRIVE, (2) STARE, (3) CHASE\_DB1, (4) HFR, and (5) ROP image.}
    \label{fig:dataset-examples}
\end{figure*}

In this section we give a brief overview of publicly available datasets for blood vessel segmentation from retinal images. We include DRIVE \cite{c1}, STARE \cite{c2} and CHASE\_DB1 \cite{c3} datasets, which are mainstream datasets for automatic vessel extraction, as well as less known HFR \cite{c4} image dataset. Example images from these four datasets and ROP image are shown in Fig. \ref{fig:dataset-examples}. Next, we introduce the reader to the recent work on automatic blood vessel segmentation using convolutional neural networks. For comprehensive review of methods for retinal vessel segmentation, including both machine learning approaches not based on neural networks and non-machine learning approaches, we direct the reader to the work of Jin et al. \cite{c5} and Oliveira et al. \cite{c6}. Here we discuss solely CNN based solutions since by quantitative measurements they have proven to be the most effective in blood vessel segmentation from a color retinal image.

\subsection{Retinal datasets for blood vessel segmentation}
The DRIVE dataset \cite{c1} includes 40 color retinal images measuring $768 \times 584$ pixels collected as a part of a Netherlands screening program for diabetic retinopathy. Images are captured using a non-mydriatic 3CCD camera at 45° FoV. All images in the dataset belong to adults of age 25 to 90, with a fully developed blood vessel network. Seven images show signs of mild vessel degradation as a result of diabetic retinopathy. All images are accompanied with manually labeled groundtruth vessel segmentation masks and FoV masks. One groundtruth set is labeled by an expert in ophthalmology, while the other is labeled by an observer trained by the ophthalmologist.

The STARE dataset \cite{c2} contains 20 color retinal images measuring $700 \times 605$ pixels extracted from the larger database created as a part of STARE project. The images are obtained using Topcon TRV-50 fundus camera at 35° FoV. Ten images show signs of different diseases that distort or obscure the blood vessel appearance, but none of the images show signs of retinopathy of prematurity. For the full diagnosis list, the reader is referred to the STARE project’s page. Remaining 10 images show no signs of disease. Similarly to DRIVE, this dataset comes with two sets of manually annotated groundtruth blood vessel masks.

CHASE\_DB1 \cite{c3} dataset contains 28 color retinal images measuring $1280 \times 960$ pixels. The images are captured by the Nidek NM-200-D handheld fundus camera with a 30° FoV. Initially, the dataset was created to validate the performance of a novel computer software for retinal vessel tortuosity measurements in school-aged children. In the dataset, there are images of left and right eyes for 14 different children, each image with two sets of groundtruth blood vessel masks. Since tortuosity is one of the key factors in diagnosis of retinopathy of prematurity, it is expected that networks trained on this dataset could perform better on ROP images. 

The HFR dataset \cite{c4} includes 45 color retinal images of size $1280 \times 960$ pixels. The images are captured using a Canon CF-60UVi camera. The dataset includes 15 images of healthy patients, 15 glaucoma images and 15 images showing diabetic retinopathy. Images are accompanied with two manually labeled groundtruth blood vessel masks, as well as FoV masks.

\subsection{Convolutional neural networks for blood vessel segmentation from retinal images}
The publication of UNet architecture \cite{c7} for medical image segmentation boosted development of CNN based solutions in many medical applications including retinal blood vessel segmentation \cite{c8,c5,c6,c11,c13,c14}. Confronting the need for a large dataset to train a well performing CNN, Ronneberger et al. have demonstrated that the UNet architecture can produce state-of-the-art results for image segmentation task, while being trained on a small dataset. Availability of small public datasets for retinal blood vessel segmentation directed the efforts toward CNNs based on UNet architecture. There have been many modifications of the original UNet architecture to address different problems occurring in blood vessel segmentation, such as varying vessel width, shape and orientation, low image quality, as well as vessel branching and crossings which often result in misclassification of a single vessel as two separate vessels. In \cite{c8} authors address the appearance of thin false vessels by incorporating UNet as a generator of a Generative Adverserial Network (GAN). Jan et al. \cite{c5} proposed the network inspired by UNet and Deformable convolutional networks \cite{c9} to address variability in shapes and scales of blood vessels. They replace the convolutional layer with a deformable convolutional block resulting in adaptive adjustment of a receptive field to the vessels’ scales and shapes during network training. Similar issues have been addressed in \cite{c6} where authors cope with vessels’ scale and orientation variation by calculating Stationary Wavelet Transform (SWT) and using it as additional information about vessels in network training. In \cite{c11} authors propose an addition to a CNN named Graph Convolutional Network \cite{c12} (GCN). While the CNN filters learn local per-pixel features, the GCN learns the global structure of the vessel network using the fact that the vessel network can be represented as a graph. The final prediction is derived by combining CNN and GCN learned features. Another approach to improve segmentation is to chain multiple networks as demonstrated in \cite{c13, c14}. The author of the LadderNet \cite{c13} propose UNet modification where multiple encoders and decoders with skip connections are chained to improve data flow, resulting in a network able to better learn features from the input image. Similar idea is employed with IterNet \cite{c14} where multiple networks are chained to iteratively improve the initial prediction. The first network in a chain is the UNet which produces the initial prediction for the given input color image. The output of the second last layer is then iteratively refined by lightweight UNets and the final prediction is the output of the last network in a chain.

All mentioned solutions use DRIVE in test and training stages. Very often, STARE \cite{c8} or CHASE\_DB1 \cite{c13} datasets, or both \cite{c5,c6,c14}, are used alongside DRIVE to create and evaluate the proposed network. The exception is \cite{c5} where WIDE \cite{c17} and SYNTHE \cite{c18} datasets are used to additionally test the proposed network. These datasets are not often used since they do not have groundtruth labels and the second dataset is synthetically generated. To measure the network performance, authors often report network’s prediction accuracy \cite{c5,c6,c13,c14}, sensitivity \cite{c5,c6,c13,c14}, specificity \cite{c5,c6,c13,c14} and precision \cite{c8,c14}, as well as derived measures. To be comparable to previously acquired results, we also adopt these measures in our study. 

\section{Methodology}\label{sec:methodology}
In this section we report details of the dataset used for evaluation, experimental protocol, and measures used for network performance evaluation.

\begin{figure*}
	\centering
	\includegraphics[ height=\textheight]{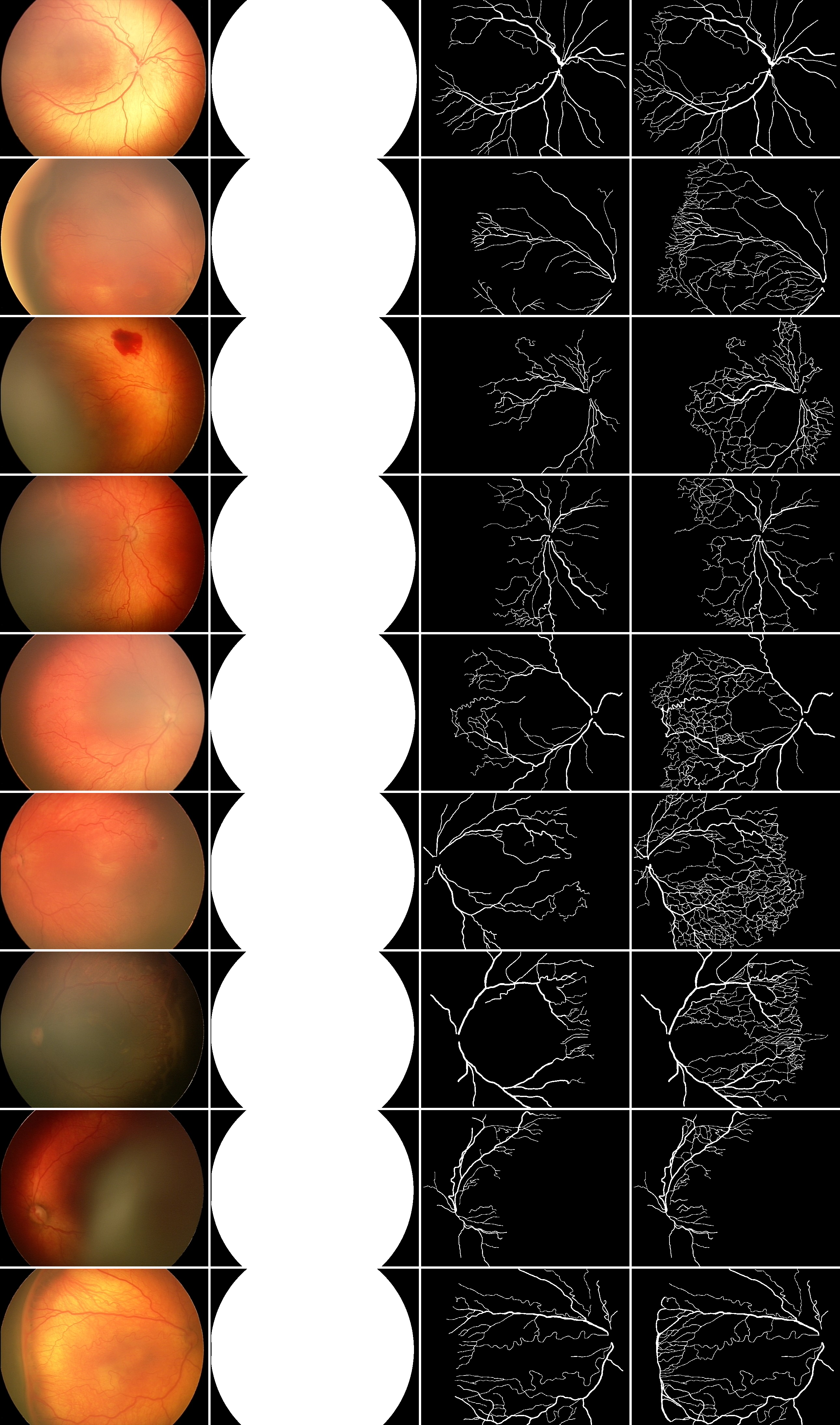}
	\caption{Images used for the CNN evaluation. From left to right: (1) a color image showing sings of ROP, (2) FoV mask, (3) the first groundtruth, (4) the second groundtruth.}
	\label{fig:dataset}
\end{figure*}

\subsection{The dataset}\label{sec:dataset}
The dataset used in this experiment consists of 9 color retinal images shown in the first column of Fig. \ref{fig:dataset}. To capture the images, a wide-field RetCam3 fundus camera with 130° FoV was used. Resulting images are of size $480 \times 640$ and are saved in JPEG file format. Abnormal vessel tortuosity and dilatation are present on all images. To test the networks on complex vessel patterns, we include images with visible excessive vessel growth and presence of vessel loops (e.g. rows 3, 5, 6 and 9 in Fig. \ref{fig:dataset}). For each color image, we provide two groundtruth blood vessel masks (columns 3 and 4 in Fig. \ref{fig:dataset}) and FoV mask (the second column in Fig. \ref{fig:dataset}). The groundtruth masks are labeled by two different observers. The first groundtruth is labeled by a PhD student in computing and control engineering which was previously trained by the experienced ophthalmologist. This is not a new approach in blood vessel labeling, and it has already been employed to create groundtruth labels for widely accepted datasets such as DRIVE and STARE. The second groundtruth is labeled by an expert ophthalmologist with more than 20 years of experience in ROP screening. The expert refined the coarse groundtruth masks created by the first observer to produce the second groundtruth. To achieve maximum accuracy, FoV masks are also labeled manually. Since groundtruth labeling is time consuming and requires expert involvement, we limit ourselves in this experiment to 9 images which are used solely for testing purposes.  

\subsection{Experimental protocol}\label{sec:experimental_protocol}
We feed color retinal images from the ROP image dataset to the neural network models proposed by Li et al. \cite{c14}, Zhuang \cite{c13} and Oliveira et al. \cite{c6}. These networks are considered to achieve state-of-the-art results in blood vessel segmentation from retinal images in terms of accuracy, sensitivity and specificity, as can be seen in Table \ref{tabl:drive-stare-chase-performance}. For each network architecture we use the latest pretrained model provided by authors. All pretrained models are made open by authors and can be downloaded from GitHub repositories following the work of Li et al., Oliveira et al. and Zhuang. Each network outputs a probability map of the same size as an input image. Values in the probability map range from 0 to 1 indicating probability that the corresponding pixel on the input map is a part of the blood vessel. A probability map is then thresholded with 0.5 threshold value to obtain a binary segmentation mask where white pixels indicate that the corresponding input pixel is a part of the blood vessel, whereas the black pixel indicates opposite. All experiments are performed on a personal computer with 32GBs of RAM, Ryzen 3700X central processing unit and NVIDIA RTX2060S graphics card.

\subsection{Performance Evaluation Metrics}
To evaluate models performance, we adopt commonly used metrics in evaluation of blood vessel segmentation solutions. These metrics include Accuracy (ACC), True Positive Rate (TPR), True Negative Rate (TNR), F-measure ($F_{1}$), and Area Under Curve (AUC) of a Receiver Operating Characteristics (ROC) graph. The accuracy measures ratio of image pixels correctly classified as a blood vessel or a background pixel. True Positive Rate and True Negative Rate, also known as sensitivity and specificity, are more specific and measure ratio of pixels correctly classified as a blood vessel or a background pixel, respectively. F-measure is used as an integral measure of the model’s precision and specificity.

\section{Results and Discussion}\label{sec:results-and-discussion}
Three models are compared in terms of accuracy, sensitivity, specificity, F-measure and AUC on the images from our dataset as shown in Table \ref{tabl:drive-stare-chase-performance}. All values are calculated on predicted segmentation masks with applied FoV masks. In Table \ref{tabl:drive-stare-chase-performance} we also report numerical results from corresponding papers for DRIVE, STARE, and CHASE\_DB1 datasets to compare them to our results. We evaluate models on two sets of groundtruth segmentation masks to evaluate the models in predicting thicker, easier to see blood vessels annotated by non-expert against small and hard to see blood vessels annotated by the expert.

\begin{table}
	\caption{Performance on DRIVE, STARE, CHASE\_DB1 and our dataset (GT1 - first groundtruth, GT2 - second groundtruth)}
	\centering
	\begin{tabular}{p{10mm}cccccc}
		\hline
		& & \multicolumn{5}{c}{Datasets}\\
		\cline{3-7}
		& & ACC & TPR & TNR & F1 & AUC \\
		\hline
		\hline
		\multirow{5}{*}{Li et al.}
		& DRIVE & 0.9574 & 0.7791 & 0.9831 & 0.8218 & 0.9816 \\
		& STARE & 0.9782 & 0.7715 & 0.9919 & 0.8146 & 0.9915 \\
		& CHASE & 0.9760 & 0.7969 & 0.9881 & 0.8073 & 0.9899 \\
		& GT1 & 0.9678 & 0.4990 & 0.9881 & 0.5513 & 0.9378 \\
		& GT2 & 0.9429 & 0.3457 & 0.9884 & 0.4307 & 0.8396 \\
		\hline
		\multirow{5}{*}{\makecell{Oliveira\\ et al.}}
		& DRIVE & 0.9576 & 0.8039 & 0.9804 & - & 0.9821 \\
		& STARE & 0.9694 & 0.8315 & 0.9858 & - & 0.9905 \\
		& CHASE & 0.9653 & 0.7779 & 0.9864 & - & 0.9855 \\
		& GT1 & 0.9681 & 0.5271 & 0.9871 & 0.5685 & 0.9125 \\
		& GT2 & 0.9436 & 0.3622 & 0.9875 & 0.4459 & 0.8250 \\
		\hline
		\multirow{5}{*}{Zhuang}
		& DRIVE & 0.9561 & 0.7856 & 0.9810 & 0.8202 & 0.9793 \\
		& STARE & - & - & - & - & - \\
		& CHASE & 0.9656 & 0.7978 & 0.9818 & 0.8031 & 0.9855\\
		& GT1 & 0.9639 & 0.5244 & 0.9830 & 0.5384 & 0.9307 \\
		& GT2 & 0.9393 & 0.3633 & 0.9833 & 0.4291 & 0.8211 \\
		\hline
	\end{tabular}
	\label{tabl:drive-stare-chase-performance}
\end{table}

\begin{figure}[!t]
	\centering
	\includegraphics[width=\linewidth]{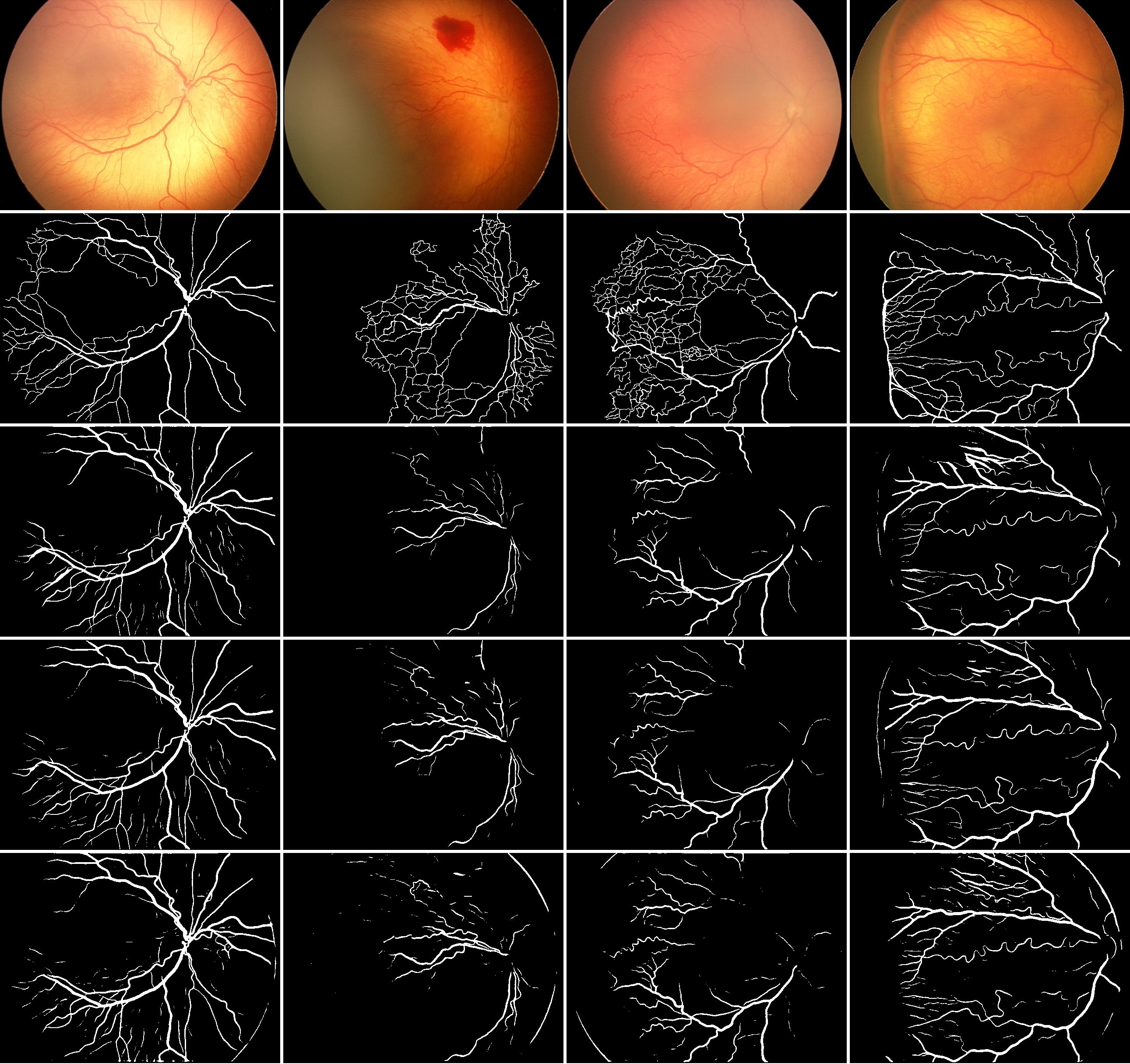}
	\caption{Predicted segmentation maps. From top to bottom rows are shown: (1) original images, (2) expert’s groundtruth and segmentation maps produced by the methods of (3) Li et al., (4) Oliveira et al., and (5) Zhuang.}
	\label{fig:results}
\end{figure}

When compared to the first set of groundtruth values, all three models achieve above 96\% accuracy and above 98\% specificity which is comparable to the accuracy and specificity values reported on publicly available datasets. However, models’ sensitivity significantly drops from approximately 77\% for public datasets to approximately 50\% for our dataset, as the models misclassified more blood vessel pixels as background pixels. Degradation in model’s performance is also visible in AUC which is reduced up to 7.9\% indicating that models produce less reliable predictions on the ROP images. The predicted segmentation masks are shown in Fig. \ref{fig:results}. It can be seen that even thick blood vessels are not segmented correctly if illumination issues are present on the image which is often the case in infant retinal images due to wide-field lenses used for screening.

Comparison with the second set of groundtruth values yields further degradation in terms of segmentation accuracy, sensitivity and AUC, which drop for approximately 2\%, 15\% and 15\% compared to the performance calculated on the first groundtruth dataset. The aggressive reduction in sensitivity and AUC is caused by the inability of the existing models to segment tiny blood vessels labeled by the expert as it can be seen in Fig. \ref{fig:results}.

Existing solutions for retinal blood vessel segmentation also produce noise in predicted segmentation maps, due to inability to differ between choroidal and retinal blood vessels. Despite being blood vessels, choroidal blood vessels are not retinal blood vessels and do not play part in ROP diagnosis. Segmentation of these vessels alongside retinal vessels would aggravate ROP diagnosis based on segmented vessels instead of alleviating it. Since the chosen models perform outstandingly on public datasets, it is our educated assumption that issues with choroidal blood vessels are caused by the lack of images showing these vessels in public datasets used for network training.

\section{Conclusions}\label{sec:conclusions}
In this paper we evaluated three state-of-the-art solutions for blood vessel segmentation from color retinal images in context of retinopathy of prematurity. The chosen solutions perform outstandingly when tested on publicly available datasets. In this study we created a dataset consisting solely of ROP images to test how well the solutions perform on ROP images and can they be used as-is for blood vessel segmentation from ROP images. Numerical measurements and visual results show that all three solutions perform similarly and that all three have significant issues with uneven illumination, infant’s thin blood vessels and choroidal blood vessels present on images. There are two possible reasons for segmentation performance degradation: (1) the images from the few public datasets that are used for training do not have properties needed for effective retinal blood vessel segmentation from ROP images and (2) trained models are not complex enough to successfully address issues present on ROP images. Since the models tend to under classify not only thin vessels, but in some cases also thick vessels which is not the problem observed on the public datasets, we believe that not all appearing issues can be solved by creating more complex CNN architecture and that retinal blood vessel segmentation CNNs would benefit from a new, ROP-aware retinal image dataset that would be used in CNN training. Additionally, we observe that existing models have learned to classify blood vessels, but not to distinguish between retinal and choroidal blood vessels which is possibly due to the lack of images showing choroidal blood vessels in public datasets.

As future work, more effort should be directed towards creating a more adequate dataset that could be used to train efficient CNNs for blood vessel segmentation from ROP retinal images. Alternatively, if ROP images are too complex for proposed CNN architectures, more complex architecture could be used to try to improve segmentation performance on existing datasets.

\bibliographystyle{IEEEtran} 
\bibliography{IEEEabrv,sisy2020refs}

\begin{thebibliography}{10}
\providecommand{\url}[1]{#1}
\csname url@samestyle\endcsname
\providecommand{\newblock}{\relax}
\providecommand{\bibinfo}[2]{#2}
\providecommand{\BIBentrySTDinterwordspacing}{\spaceskip=0pt\relax}
\providecommand{\BIBentryALTinterwordstretchfactor}{4}
\providecommand{\BIBentryALTinterwordspacing}{\spaceskip=\fontdimen2\font plus
\BIBentryALTinterwordstretchfactor\fontdimen3\font minus
  \fontdimen4\font\relax}
\providecommand{\BIBforeignlanguage}[2]{{%
\expandafter\ifx\csname l@#1\endcsname\relax
\typeout{** WARNING: IEEEtran.bst: No hyphenation pattern has been}%
\typeout{** loaded for the language `#1'. Using the pattern for}%
\typeout{** the default language instead.}%
\else
\language=\csname l@#1\endcsname
\fi
#2}}
\providecommand{\BIBdecl}{\relax}
\BIBdecl

\bibitem{c15}
I.~C. for the Classification of Retinopathy~of Prematurity \emph{et~al.}, ``The
  international classification of retinopathy of prematurity revisited.''
  \emph{Archives of ophthalmology (Chicago, Ill.: 1960)}, vol. 123, no.~7, p.
  991, 2005.

\bibitem{c16}
E.~Poletti, D.~Fiorin, E.~Grisan, and A.~Ruggeri, ``Automatic vessel
  segmentation in wide-field retina images of infants with retinopathy of
  prematurity,'' in \emph{2011 Annual International Conference of the IEEE
  Engineering in Medicine and Biology Society}.\hskip 1em plus 0.5em minus
  0.4em\relax IEEE, 2011, pp. 3954--3957.

\bibitem{c5}
Q.~Jin, Z.~Meng, T.~D. Pham, Q.~Chen, L.~Wei, and R.~Su, ``Dunet: A deformable
  network for retinal vessel segmentation,'' \emph{Knowledge-Based Systems},
  vol. 178, pp. 149--162, 2019.

\bibitem{c6}
A.~Oliveira, S.~Pereira, and C.~A. Silva, ``Retinal vessel segmentation based
  on fully convolutional neural networks,'' \emph{Expert Systems with
  Applications}, vol. 112, pp. 229--242, 2018.

\bibitem{c7}
O.~Ronneberger, P.~Fischer, and T.~Brox, ``U-net: Convolutional networks for
  biomedical image segmentation,'' in \emph{International Conference on Medical
  image computing and computer-assisted intervention}.\hskip 1em plus 0.5em
  minus 0.4em\relax Springer, 2015, pp. 234--241.

\bibitem{c8}
J.~Son, S.~J. Park, and K.-H. Jung, ``Retinal vessel segmentation in
  fundoscopic images with generative adversarial networks,'' \emph{arXiv
  preprint arXiv:1706.09318}, 2017.

\bibitem{c11}
S.~Y. Shin, S.~Lee, I.~D. Yun, and K.~M. Lee, ``Deep vessel segmentation by
  learning graphical connectivity,'' \emph{Medical image analysis}, vol.~58, p.
  101556, 2019.

\bibitem{c12}
T.~N. Kipf and M.~Welling, ``Semi-supervised classification with graph
  convolutional networks,'' \emph{arXiv preprint arXiv:1609.02907}, 2016.

\bibitem{c14}
L.~Li, M.~Verma, Y.~Nakashima, H.~Nagahara, and R.~Kawasaki, ``Iternet: Retinal
  image segmentation utilizing structural redundancy in vessel networks,'' in
  \emph{The IEEE Winter Conference on Applications of Computer Vision}, 2020,
  pp. 3656--3665.

\bibitem{c13}
J.~Zhuang, ``Laddernet: Multi-path networks based on u-net for medical image
  segmentation,'' \emph{arXiv preprint arXiv:1810.07810}, 2018.

\bibitem{c1}
J.~Staal, M.~D. Abr{\`a}moff, M.~Niemeijer, M.~A. Viergever, and
  B.~Van~Ginneken, ``Ridge-based vessel segmentation in color images of the
  retina,'' \emph{IEEE transactions on medical imaging}, vol.~23, no.~4, pp.
  501--509, 2004.

\bibitem{c2}
A.~Hoover, V.~Kouznetsova, and M.~Goldbaum, ``Locating blood vessels in retinal
  images by piecewise threshold probing of a matched filter response,''
  \emph{IEEE Transactions on Medical imaging}, vol.~19, no.~3, pp. 203--210,
  2000.

\bibitem{c3}
C.~G. Owen, A.~R. Rudnicka, R.~Mullen, S.~A. Barman, D.~Monekosso, P.~H.
  Whincup, J.~Ng, and C.~Paterson, ``Measuring retinal vessel tortuosity in
  10-year-old children: validation of the computer-assisted image analysis of
  the retina (caiar) program,'' \emph{Investigative ophthalmology \& visual
  science}, vol.~50, no.~5, pp. 2004--2010, 2009.

\bibitem{c4}
A.~Budai, R.~Bock, A.~Maier, J.~Hornegger, and G.~Michelson, ``Robust vessel
  segmentation in fundus images,'' \emph{International journal of biomedical
  imaging}, vol. 2013, 2013.

\bibitem{c9}
J.~Dai, H.~Qi, Y.~Xiong, Y.~Li, G.~Zhang, H.~Hu, and Y.~Wei, ``Deformable
  convolutional networks,'' in \emph{Proceedings of the IEEE international
  conference on computer vision}, 2017, pp. 764--773.

\bibitem{c17}
R.~Estrada, C.~Tomasi, S.~C. Schmidler, and S.~Farsiu, ``Tree topology
  estimation,'' \emph{IEEE transactions on pattern analysis and machine
  intelligence}, vol.~37, no.~8, pp. 1688--1701, 2014.

\bibitem{c18}
H.~Zhao, H.~Li, S.~Maurer-Stroh, and L.~Cheng, ``Synthesizing retinal and
  neuronal images with generative adversarial nets,'' \emph{Medical image
  analysis}, vol.~49, pp. 14--26, 2018.

\end{thebibliography}

\end{document}